\documentclass[12pt,preprint]{aastex}
\begin{document}

\def\be{\begin{equation}}
\def\ee{\end{equation}}
\def\lesssim{\raisebox{-0.3ex}{\mbox{$\stackrel{<}{_\sim} \,$}}}
\def\gtrsim{\raisebox{-0.3ex}{\mbox{$\stackrel{>}{_\sim} \,$}}}

\title{The intrinsic intensity modulation of PSR B1937+21 at 1410 MHz}

\author{Fredrick A. Jenet \altaffilmark{1}, Janusz Gil\altaffilmark{2}}
\altaffiltext{1}{California Institute of Technology, Jet Propulsion Laboratory\\ 4800 Oak Grove Drive,Pasadena, CA
91109} \altaffiltext{2}{Institute of Astronomy, University of Zibeline G\'ora \\Lubuska 2, 65-265, Zibeline
G\'ora, Poland}

\begin{abstract}
The single-pulse properties of the millisecond radio pulsar PSR
B1937+21 are studied in the 1410 MHz radio band. Aside from occasional
``giant pulses'' occurring in restricted regions of pulse phase, the
emission appears to be remarkably stable, showing no pulse-to-pulse
fluctuations other then those induced by propagation through the
interstellar medium. This type of behavior has not been seen in any
other pulsar although it was seen in previous 430 MHz observations of
this source. 

The stability of PSR B1937+21 can be understood in the context of the
sparking gap model of radio pulsar emission. Given the emission
properties of this source at 430 MHz, this model predicts that the
emission at all higher frequencies will be just as stable. Since the
stability depends on the outflow velocity of the emitting plasma, an
upper bound may be placed on its Lorentz factor.

\end{abstract}
\section{Introduction}

This letter reports on observations of the radio pulsar PSR B1937+21
taken with a center frequency of 1410 MHz at the 305m Arecibo radio
telescope in Puerto Rico. Previous observations of this source at 430
MHz revealed a behavior that was completely different from other known
sources \citep[]{jenal01}. At 430 MHz this source
exhibits almost no detectable pulse-to-pulse fluctuations. Occasional bursts
of radio radiation, or ``giant pulses'', are observed but they are
restricted to small regions in pulse phase \citep{kt00, cst+96}. The
observations presented here show that this remarkable pulse-to-pulse
stability of the non-giant pulse emission also occurs at 1410 MHz.
\citet{es03} attempted to measure the modulation properties of PSR
B1937+21 at 1300 MHz with the Westerbork synthesis radio
telescope. Unfortunately, the signal-to-noise ratio in their
observations was not high enough to measure the pulse-to-pulse
properties.

Observations of bright pulsars have shown that the shapes and
intensities of individual pulses are unique, although they average
together to form a stable mean profile. The characteristic widths of
individual pulses, typically referred to as sub-pulses, are usually
smaller than the average profile width. This typical behavior has not been
detected in the pulses from PSR B1937+21. Instead, this source appears
to be highly stable in the sense that each pulse is identical to the
next.








It will be shown that the stability of this source at both frequencies
can be understood within the context of the sparking gap model of
pulsar emission originally proposed by \citet{rs75} and later reworked
in more detail by \citet[][ hereafter GS00] {gs00}. In this model, the
stability is a result of a large number of emission regions, or
``sparks'', occurring on the polar cap. The GS00 model predicts that if
the emission is stable at a given frequency, then it will be stable at
all higher frequencies. Since this stability will also depend on the
velocity of the emitting plasma, an upper bound may be placed on its Lorentz factor.

The sparking-gap model of \citet{rs75} and its revised form (GS00)
have received much attention in recent years.  They have been used to
interpret the sub-pulse properties of slow pulsar conal emission
\citep{esv03,rr03,vsr+03,ad01,dr01} and of millisecond pulsars core
emission \citep{es03}. These models have also been used in pulsar
population studies \citep{acc02,fcm01}. Given their widespread use, it
is important to develop observational tests which can determine the
validity of these models. Since the GS00 model predicts that the radio
emission from PSR B1937+21 should be stable at all frequencies higher
than 430 MHz, high frequency observations of this source will test
the current form of the GS00 model.


The next section describes the 1410 MHz observations performed at the
Arecibo radio observatory. It also describes the techniques used to
measure the intrinsic fluctuation properties of this source. Section 3
discusses the sparking gap model in more detail with emphasis on the
frequency dependence of the modulation index. This letter is
summarized in section 4.


\section{Observations and Analysis}
The data were taken at the 305 m Arecibo radio telescope using the
1400 MHz Gregorian dome receiver. The observation lasted two hours and
approximately 4.5 million pulses were obtained. The exact center
frequency used was 1410 MHz. Both circular polarizations were two-bit
complex sampled at a rate of 10 MHz (100 ns) and recorded to tape
using the Caltech baseband recorder \citep{jenal98}. Further
processing of the data were performed at the Caltech Center for
Advanced Computation and Research (CACR) using a 256 processor
Hewlett-Packard Exemplar. The two-bit complex samples were unpacked
and assigned optimum values in order to minimize signal distortion
\citep{ja98}. The dual polarization voltage data was adjusted using an
empirically derived cross-talk matrix \citep{sti82}. The effect of the
Earth's motion around the Sun was removed by resampling the complex
voltage data at a rate necessary to transform the data into the
barycentric frame. This rate was calculated using the software package
TEMPO\footnote{http://pulsar.princeton.edu/tempo}. The effects of
interstellar dispersion were removed by coherently dedispersing the
data \citep{jenal98,hr75} at a dispersion measure of 71.0249 pc
$\mbox{cm}^3$ \citep{jenal01}.

As with the 430 MHz observations, the single pulses from PSR B1937+21
at 1410 MHz are not bright enough to use conventional single pulse
analysis techniques. Hence, the low intensity analysis methods used
by \citet{es03} and \citet{jenal01} are used here. Typically, a pulsar
exhibits pulse-to-pulse shape variability and intensity
fluctuations. Three statistical techniques are used to determine if
these phenomena are occurring in this source.  First, the phase
resolved modulation index is calculated in order to determine if any
pulse-to-pulse amplitude fluctuations are present. Second, the average
intensity fluctuation spectrum \citep{bac73} is calculated to
determine the time scale of the modulation. Third, the average
autocorrelation function (ACF) and the ACF of the average profile are
calculated and compared to each other in order to determine if there
are any pulse-to-pulse shape variations occurring. Pulse-to-pulse
fluctuations may be intrinsic to the source or due to propagation
through the ISM. The magnitude of the modulation index, the time scale
of the modulation, and the existence of any pulse shape variations can
be used to discriminate between intrinsic and ISM induced
pulse-to-pulse variations.

\begin{figure}
\caption{\label{figure3}Average Pulse Profile of PSR B1937+21}
\plotone{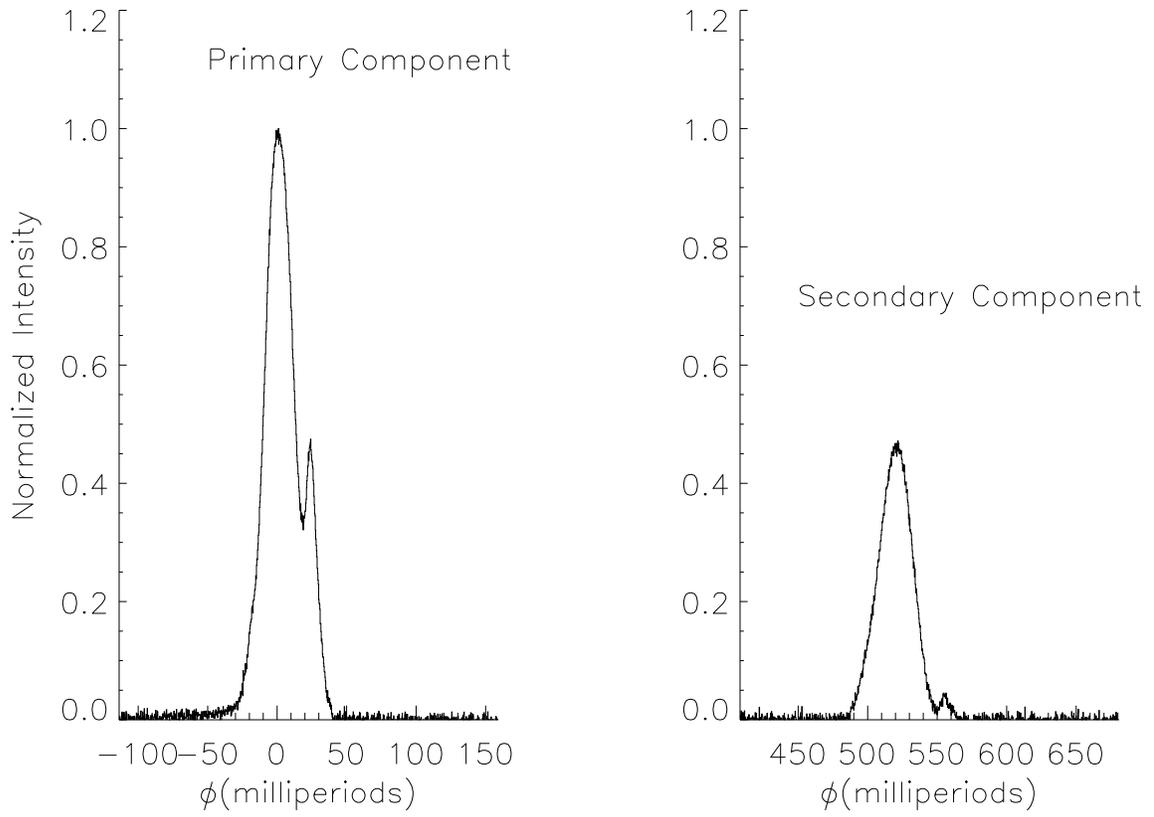}
\end{figure}
\begin{figure}
\caption{\label{figure1}Autocorrelation functions of the primary and secondary components.}


\plotone{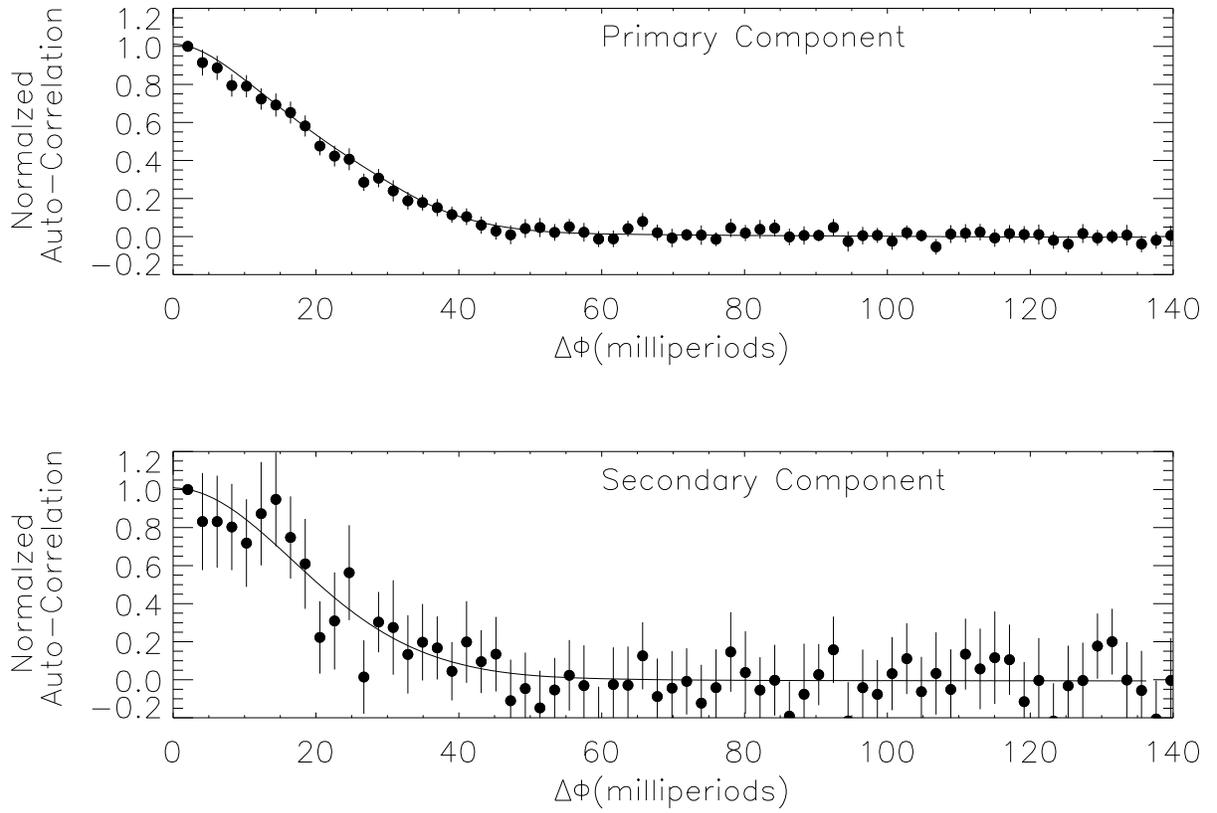}
\end{figure}

The average pulse profile, $<I_s(\phi)>$, normalized by its peak value
is shown in figure \ref{figure3}. $I_s(\phi)$ is the signal intensity
at pulse phase $\phi$ and the angle brackets, $<>$, represent averaging
over the ensemble of 4.5 million single pulses. The average profile
consists of two components separated by about 520 milli-periods
(mP). Note that only a small temporal region of length 425 $\mu$s (273
mP) is shown around each component with a time resolution of .42 $\mu
s$. The giant pulses occur between 35 and 40 mP near the primary
component and between 560 and 565 mP in the secondary component. The
autocorrelation function of the average profile, $C_{<I_s>}(\Delta
\phi)$, is calculated for each component separately and compared to
the average autocorrelation function, $<C_{I_s}(\Delta \phi)>$, of
each component. These correlation functions are defined as:
\begin{eqnarray}
C_{<I_s>}(\Delta \phi) &=& \frac{1}{\phi_1 - \phi_0} \int_{\phi_0}^{\phi_1} <I_s(\phi)><I_s(\phi + \Delta \phi)> d\phi \\
<C_{I_s}>(\Delta \phi) &=& \frac{1}{\phi_1 - \phi_0}<\int_{\phi_0}^{\phi_1} I_s(\phi)I_s(\phi + \Delta \phi) d\phi >,
\end{eqnarray}
where $\phi_0$ and $\phi_1$ are the starting and ending pulse phases of
the given emission component, respectively, and $\Delta \phi$ is the
phase lag. Note that cyclic boundary conditions are assumed when
evaluating the above expressions for the case when $\phi + \Delta
\phi$ lies outside the interval $[\phi_0,\phi_1]$.  If the individual
pulses have identical shapes from pulse-to-pulse, then these two
autocorrelation function will have identical shapes. Figure
\ref{figure1} shows both autocorrelation functions plotted together for each
component. When calculating the average ACF, $<C_{I_s}>$, the high
resolution ACF was rebined to a time resolution of 3.2 $\mu s$ (2.05 mP)
in order to increase the signal-to-noise ratio. All the of ACFs were
normalized by their value at $\Delta \phi = 2.05$ mP. This corresponds to
the second phase-lag bin of the rebined average ACF and was chosen to
avoid the noise spike occurring in the first bin of the rebined
average ACF. The solid line represents $C_{<I_s>}(\Delta \phi)$ while
the filled circles represent $<C_{I_s}(\Delta \phi)>$. Each
polarization was analyzed separately and the resulting ACFs were
averaged together. The error bars where calculated using the number of
points averaged together within each bin and the root-mean-square (RMS) of the
data within each bin. The error bars represent the $95\%$ confidence
range. Note that even though the giant pulses are included in the
ACF analysis, they will have little effect since they are very narrow
and they occur infrequently.

\begin{figure}
\caption{\label{figure2}The phase resolved modulation indices of the primary and secondary components.}
\plotone{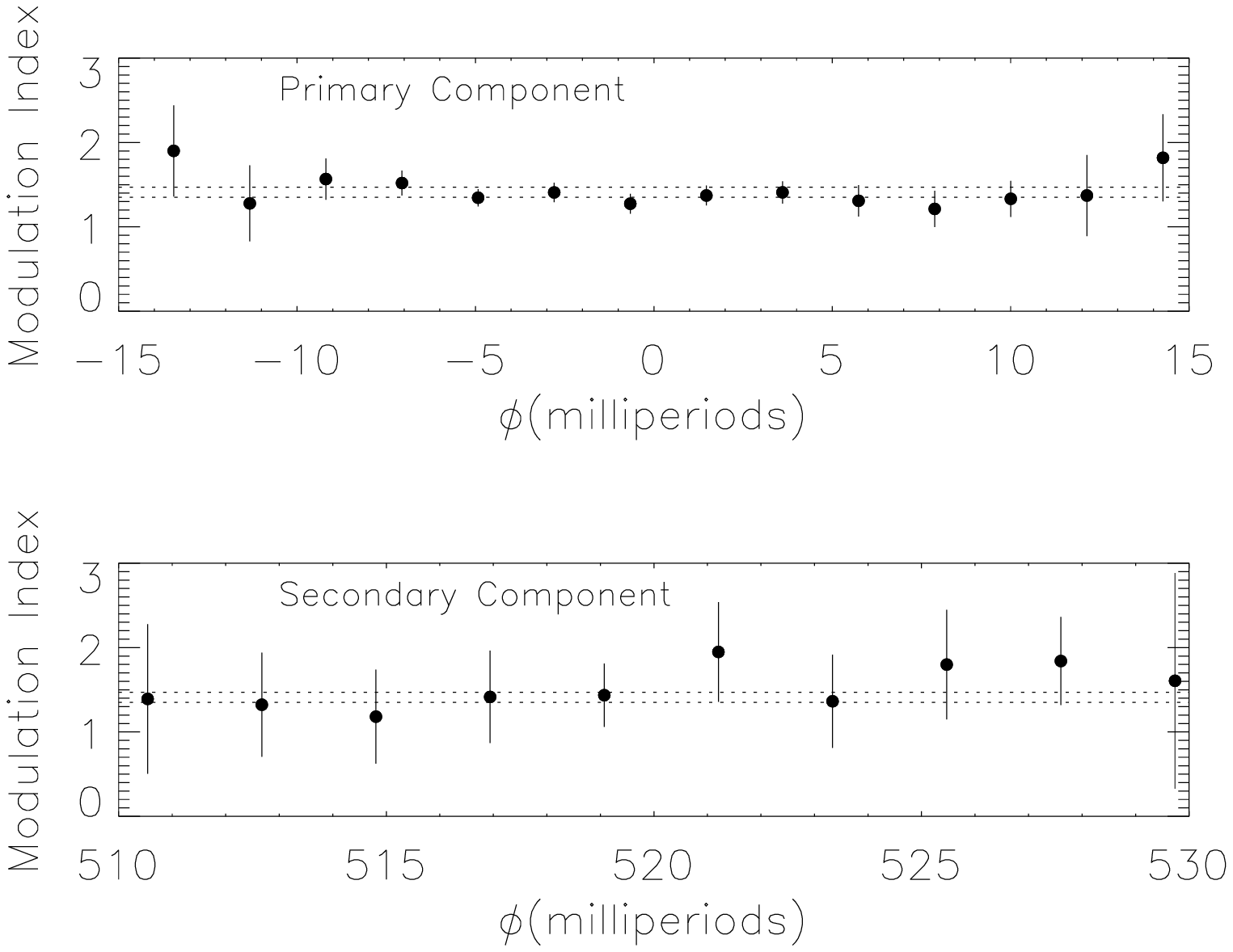}
\end{figure}

 The phase resolved modulation index is defined as
\begin{equation}
m(\phi) =\frac{ \sqrt{<I_s(\phi)^2> -<I_s(\phi)>^2}}{<I_s(\phi)>}.
\end{equation}
$m(\phi)$ is a measure of the pulse-to-pulse variation in
intensity. Again, both polarization were analyzed separately and the
resulting modulation indices were averaged together. In order to
increase the signal-to-noise ratio, both $<I_s(\phi)>$ and
$<I_s(\phi)^2>$ were rebined to a time resolution of 3.3 $\mu s$ (2.1
mP) before calculating $m(\phi)$. The error bars where calculated
using the number of points averaged within each bin and the RMS values
of $<I_s(\phi)>$ and $<I_s(\phi)^2>$ within each bin. The error bars
represent the $95\%$ confidence range. If the amplitude of each pulse
is constant and the only remaining fluctuations are due to a Gaussian
statistical process, then $m(\phi)$ will be equal to 1\footnote{Note that this
definition of $m$ is slightly different from the typical definition
which averages down the noise before calculating $m$. In that case, $m
=0$ if no pulse-to-pulse fluctuations were present.}. Figure
\ref{figure2} plots $m(\phi)$ for both components. Only the phase
region where a significant measurement could be made is shown. The
modulation index is consistent with being constant across each
component as expected if all pulses have the same shape. The average
value of the modulation index is 1.37 $\pm 0.06$ and 1.54 $\pm 0.2$
for the main and secondary components, respectively.

Since pulse-to-pulse modulation is detected, it is necessary to determine if it is intrinsic to the pulsar or a
result of propagation through the ISM. The expected modulation index due to the ISM alone is given by
\citep{jenal01}:
\begin{equation}
m = \sqrt{1 + \frac{2}{N}},
\end{equation}
where $N$ is the number of scintills in the band. $N$ may be estimated by \citep{cwd+90}:
\begin{equation}
\label{sism}
N = 1 + \eta \frac{B}{\delta \nu},
\end{equation}
where $B$ is the observing bandwidth, $\delta \nu$ is the
decorrelation bandwidth and $\eta$ is a packing fraction which lies
between 0.1 and 0.2. Using a decorrelation bandwidth of 1.1 MHz at
1410 \citep{cor86}, m is expected to be within 1.38 and 1.48. The two
dashed lines in Figure \ref{figure2} represent these two values. The
time scale of the fluctuations were determined by calculating the
average intensity fluctuation spectrum for both pulse components. Each
spectrum is consistent with zero power down to a frequency of $0.02$
$s^{-1}$. At this point, the spectrum starts to rise. This type of
spectrum is indicative of the short time scale fluctuations induced by
the ISM. The characteristic time scale as determined by that frequency
where the power falls to half its maximum value is given by 125 s,
consistent with that measured by \citet{cor86}. Since the magnitude
and time scale of the observed fluctuations are consistent with that
expected for the ISM, it is concluded that these fluctuations are
primarily due to ISM propagation and not due to intrinsic
pulse-to-pulse fluctuations.



\section{Intensity modulation and the sparking gap model}

These results can be understood in the context of the GS00 model. In
this model, a region of low charged particle density forms just above
the surface of the polar cap. This ``vacuum gap'' is postulated to
contain a number of localized plasma discharges or ``sparks''. The
plasma generated in each spark moves upwards following the local
magnetic field. At some point, processes within the spark associated
plasma generate the observed radiation with higher frequencies being
emitted at lower altitudes \citep{mgp00,glm03}. The minimum distance
between two adjacent sparks on the pulsar's surface is approximately
given by the height of the vacuum gap, $h$ (see GS00 for
details). Assuming that the polar cap is populated as densely as
possible by these sparks, the total number of sparks on the cap is
estimated by $a^2$ where $a$ is given by
\begin{equation}
a=\frac{r_p}{h} \label{a},
\end{equation}
$r_p=10^4R_6^{1.5}P^{-0.5}$ cm is the radius of the polar cap, 
$R_6$ is the neutron star radius in units of $10^6$ cm,  and $P$ is
the pulsar period in seconds.
$a$ is called the complexity parameter since it was shown by GS00 that
$a$ is correlated with the profile morphology described by \citet{r83}. Conal single
profiles have the lowest values of $a$, followed by conal doubles and
multiples and then by triple and core singles. Since $a$ is
proportional to the number of sparks that would emit into the
observers line of site, the pulse-to-pulse modulation is expected to
be anticorrelated with $a$. Possible evidence for this correlation is
discussed in \citep{jg03}. PSR B1937+21 has the highest complexity
parameter of all pulsars whose single pulse properties are
known. Hence, its stability is a result of the large number of sparks
occurring on its polar cap.

A large number of sparks on the surface is not quite enough to
guarantee that the individual pulses will be stable. The emission from
each spark-associated plasma column must overlap each other in order to
average out the fluctuations. Since the spark plasma is moving out
along the magnetic field at relativistic speeds, the angular width of
the radiation pattern, $\Theta_r$, will be given by
\begin{equation}
\Theta_r = \frac{1}{\gamma},
\label{tr}
\end{equation}
with $\gamma = 1/\sqrt{1 - (v/c)^2}$, $v$ is the velocity of the spark
associated plasma in the radio emission region, and $c$ is the speed
of light. As the spark plasma moves up along the magnetic field lines,
its radiation will be beamed in a direction tangent to the local
magnetic field line. Hence, the radiation from two sparks that
originate near each other on the polar cap will be emitted into
different directions. For a dipolar field, it turns out that the
angle, $\Delta \theta$, between the radiation beams from two adjacent
sparks will be given by
\begin{equation}
\Delta \theta = \frac{0.015}{R_6}\frac{\Delta d}{r_p}r_6^{\frac{1}{2}}P^{-\frac{1}{2}}, \label{dth}
\end{equation}
where $\Delta d$ is the distance between adjacent sparks on the polar
cap (see GS00 for details), $r_6$ is the emission altitude normalized
by the neutron star radius $R=R_610^6$
cm. In order for the emission from adjacent sparks to overlap, the
following condition must be met
\begin{equation}
\Delta \theta < \Theta_r.
\label{teq}
\end{equation}
When the above equation is satisfied, the emission from adjacent
sparks will overlap and the corresponding radio emission will show
very little pulse-to-pulse intensity fluctuations provided that $a$ is
large. From Equations \ref{a}, \ref{tr}, and \ref{dth} one can rewrite
equation \ref{teq} as
\begin{equation}
\gamma < \frac{a R_6}{0.015}\left(\frac{P}{r_6}\right)^{\frac{1}{2}},
\label{gamma2}
\end{equation}
where $\Delta d$ was taken to be equal to $h$. If
the emission from a given altitude satisfies the above condition, then
emission from lower altitudes will also satisfy this condition. Hence,
if emission from a given altitude is observed to be stable, then
emission from all lower altitudes will also be stable. Observations of
other pulsars point to the existence of a ``radius-to-frequency''
mapping which states that higher frequencies are emitted at lower
altitudes \citep[e.g.][]{kg03}. Assuming that this holds for PSR
B1937+21, the pulses emitted at 1410 MHz are expected to be at least
as stable as those emitted at 430 MHz.

Since the observations are in agreement with the sparking gap model
described above, one can go one step further and use Equation
\ref{gamma2} to place an upper bound,$\gamma_{max}$, on the Lorentz
factor of the emitting plasma. The following form of the radius-to-frequency
map will be used \citep{kg03}:
\begin{equation}
r_6(\nu_{GHz}) = 40 \nu_{GHz}^{-0.26} \dot{P}_{-15}^{0.07} P^{0.3},
\end{equation}
where $\nu_{GHz}$ is the observing frequency in GHz and
$\dot{P}_{-15}$ is the pulsar period derivative in $10^{-15}$ s/s. For
PSR B1937+21, $r_6(0.43) = 3.8$ and $a = 23$. Hence, Equation
\ref{gamma2} yields $\gamma_{\mbox{max}} = 31R_6$. This estimate
assumes the standard vacuum gap model described by RS75. For the case
of the near threshold vacuum gap (NTVG) model \citep{gm02}, $a$ can be
up to about 3 times higher then that estimated from equation \ref{a} since
$a$ will
depend on the actual value of the surface magnetic field line radius of curvature. In this case, $\gamma_{\mbox{max}} = 91R_6$.

\section{Summary and Discussion}

Radio observations of the millisecond pulsar PSR B1937+21 were taken
at a center frequency of 1410 MHz. Statistical techniques were used to
determine if this source exhibits any of the standard single pulse
phenomenology that has been seen in other radio pulsars. Remarkably,
none of the usual phenomena were detected. The data are consistent
with each pulse having the same shape and intensity. Previously,
\citet{jenal01} showed that this source exhibits the same single pulse
stability at 430 MHz. This stability is currently unique to this
source. Hence, PSR B1937+21 maybe the prototype for a new class of
radio pulsars. On the other hand, the sparking gap model provides a
unified framework that can explain this source's behaviour as an
extreme case of the normal pulsar emission process. In the context of
the GS00 model, the stability is a result of a large number of sparks
occurring on the polar cap together with a low Lorentz factor of the
emitting plasma. Since the 430 MHz observations show no pulse-to-pulse
fluctuations, the GS00 model predicts that all higher frequencies will
be just as stable provided higher frequencies are emitted at lower
altitudes. This result is independent of the exact physical model used
to describe the vacuum gap. Future observations of this source at
several frequencies above 430 MHz will further test the GS00 model. If
fluctuations are seen, then the current form of the sparking gap model
is incorrect. Observations of this source at frequencies below 430 MHz
will enable a better measurement of the Lorentz factor. If
pulse-to-pulse fluctuations are seen at lower frequencies, a lower
bound may be placed on $\gamma$ for a given model of the vacuum gap.

\acknowledgements 

Part of this research was performed at the Jet Propulsion Laboratory,
California Institute of Technology, under contract with the National
Aeronautics and Space Administration. GJ acknowledges the support of
the Polish State Committee for scientific research under grant 2 P03D 008 19.

{}
\end{document}